\documentclass[nofootinbib,twocolumn,aps,pra]{revtex4}
\usepackage{graphicx}
\usepackage{epstopdf}

\begin{document}
\title{Two band model for coherent excitonic condensates}
\author{V. Apinyan}
\author{T. K. Kope\'{c}\footnote{Tel.:  +48 71 395 4286; E-mail address: kopec@int.pan.wroc.pl.}}
\affiliation{Institute for Low Temperature and Structure Research, Polish Academy of Sciences\\
PO. Box 1410, 50-950 Wroc\l{}aw 2, Poland \\}

\begin{abstract}
%
We consider the excitonic correlations in the two band solid state system composed of the valence band and conduction band electrons. We treat the phase coherence mechanism in the system by presenting the electron operator as a fermion attached to the U(1) phase-flux tube. The emergent bosonic gauge field, related to the phase variables appears to be crucial for the coherent Bose-Einstein condensation (BEC) of excitons. We calculate the normal excitonic Green functions, and the single-particle density of states functions being a convolution between bosonic and fermionic counterparts. We obtain the total density of states (DOS) as a sum of two independent parts. For the coherent normal fermionic DOS, there is no hybridization-gap found in the system due to strong coherence effets and phase stiffness. 
\end{abstract}

\maketitle

\section{\label{sec:Section_1} Introduction}
%
The excitonic quasiparticles, in the solid state materials lead to a very rich and interesting physical phenomena, one of which is the excitonic insulator (EI) state predicted many years ago by Keldysh and Kozlov  \cite{cite-1,cite-2, cite-3}. Anoher fascinating phenomenon that should be mentioned is the Bose-Einstein-Condensation (BEC) of excitons at cryogenic temperatures. Despite many experimental efforts to obtain the coherent excitonic condensates \cite{cite-4, cite-5,cite-6,cite-7,cite-8} there is not yet a definitive evidence for such states. The low density system of excitons behaves like the usual Bose gas, while the high density system of bound e-h pairs behaves like the system of weakly coupled Cooper pairs. Thus the resulting BCS-BEC crossover \cite{cite-9,cite-10,cite-11,cite-12,cite-13,cite-14} represents an interesting theoretical problem typical for the excitonic systems. 

The importance of the phase coherence in the excitonic pair plasma is discussed recently in Refs.\onlinecite{cite-9,cite-15} and also in Ref.\onlinecite{cite-17}. The author, in Refs.\onlinecite{cite-15,cite-16}, shows from general considerations that the coherent BEC transition critical temperature should differs from the excitonic pair formation critical temperature. More stronger theoretical demonstration of this fact is given in Ref.\onlinecite{cite-9}, where it is shown that the excitonic insulator state is an excitonium state, where the \textit{incoherent} e-h bound pairs are formed, and furthermore, at lower temperatures, the BEC of excitons appears in consequence of reconfiguration and \textit{coherent} condensation of the preformed excitonic pairs. In the whole BCS-BEC transition region the e-h mass difference leads to a large suppression of the BEC transition temperature, which is proved to be not the same as the excitonic pair formation temperature \cite{cite-9}.  

The particle coherence in the usual sense of hybridization between the conduction band electron and valence band holes is discussed in many works \cite{cite-10,cite-11,cite-12,cite-13,cite-14} within the three dimensional (3D) extended Falicov-Kimball model (EFKM) with a dispersive $f$-band electrons at half-filling. It is shown recently \cite{cite-17} that the EI state is unstable when the case of pure Falikov-Kimball model (FKM) \cite{cite-18} (with fully localized bands) is approached.  

In the present paper we will employ the EFKM model with the $f$-band hopping mechanism to study the coherent excitonic condensation in the 3D system of correlated excitons. We show how the local and nonlocal excitonic correlations govern the EI state and the coherent excitonic condensate state respectively. We derive the EI state as a local contribution from on-site e-h interactions, while the coherent condensation of excitons occurs only when the nonlocal excitonic correlations are included and the phase stiffness is achieved in the whole system.

\section{\label{sec:Section_2} The model}
%
For the study coherent excitonic mechanism in 3D excitonic systems we have chosen two-band EFKM. The Hamiltonian of the EFKM model is given by
\begin{eqnarray}
&&{\cal{H}}=-t_{c}\sum_{\left\langle {\bf{r}},{\bf{r}}' \right\rangle}\left[\bar{c}({{\bf{r}}})c({{\bf{r}}}')+h.c.\right]-\bar{\mu}\sum_{{\bf{r}}}n({\bf{r}})-
\nonumber\\
&&-t_{f}\sum_{\left\langle {\bf{r}},{\bf{r}}' \right\rangle}\left[\bar{f}({{\bf{r}}})f({{\bf{r}}}')+h.c.\right]+\frac{\epsilon_{c}-\epsilon_{f}}{2}\sum_{{\bf{r}}}\tilde{n}({\bf{r}})+
\nonumber\\
&&+U\sum_{{\bf{r}}}\frac{1}{4}\left[n^{2}({\bf{r}})-\tilde{n}^{2}({\bf{r}})\right].
\label{Equation_1}
\end{eqnarray}
Here $\bar{f}({{\bf{r}}})$ ($\bar{c}({{\bf{r}}})$) creates an $f$ ($c$) electron at the lattice position ${\bf{r}}$, the summation $\left\langle {\bf{r}}, {\bf{r}}' \right\rangle$ runs over pairs of n.n. sites of 3D lattice. The density type short hand notations are introduced $n({\bf{r}})=n_{c}({\bf{r}})+n_{f}({\bf{r}})$ and $\tilde{n}({\bf{r}})=n_{c}({\bf{r}})-n_{f}({\bf{r}})$. Next, $t_{c}$ is the hopping amplitude for $c$-band electrons and $\epsilon_{c}$ is the corresponding on-site energy level. Similarly, $t_{f}$ is the hopping amplitude for $f$-band electrons and $\epsilon_{f}$ is the on-site energy level for $f$-orbital.
The on-site (local) Coulomb interaction $U$ in the last term of the Hamiltonian in Eq.(\ref{Equation_1}) plays the coupling role between the electrons in the $f$ and $c$ bands. The chemical potential $\bar{\mu}$ is $\bar{\mu}=\mu-\bar{\epsilon}$, where $\bar{\epsilon}=\left(\epsilon_{c}+\epsilon_{f}\right)/2$. We will use $t_{c}=1$ as the unit of energy and we fix the band parameter values $\epsilon_{c}=0$ and $\epsilon_{f}=-1$. For the $f$-band hopping amplitude $t_{f}$ we consider the values $t_{f}=-0.3$ and $t_{f}=-0.1$. Throughout the paper, we set $k_{B}=1$ and $\hbar=1$ and lattice constant $a=1$.
%
\section{\label{sec:Section_3} The excitonic insulator}
%
Employing the imaginary-time fermionic path integral techniques, we introduce the fermionic Grassmann variables ${f}({{\bf{r}}}\tau)$ and ${c}({{\bf{r}}}\tau)$ at each site ${\bf{r}}$ and for each time $\tau$, which varies in the interval $0\leq \tau \leq\beta$, where $\beta=1/T$ with $T$ being the thermodynamic temperature. The time-dependent variables ${c}({{\bf{r}}}\tau)$ and ${f}({{\bf{r}}}\tau)$ are satisfying the anti-periodic boundary conditions ${x}({{\bf{r}}}\tau)=-{x}({{\bf{r}}}\tau+\beta)$, where $x=f$ or $c$.
The grand canonical partition function of system of fermions written as a functional integral over the Grassmann fields is
\begin{eqnarray}
Z=\int\left[{\cal{D}}\bar{c}{\cal{D}}c\right]\left[{\cal{D}}\bar{f}{\cal{D}}f\right]e^{-{\cal{S}}[\bar{c},c, \bar{f},f]},
\label{Equation_2}
\end{eqnarray} 
where the action in the exponent is given in the path-integral formulation in the form
\begin{eqnarray}
{\cal{S}}[\bar{c},c, \bar{f},f]=\sum_{x=f,c}{\cal{S}}_{B}[\bar{x},x]+\int^{\beta}_{0}d\tau {{H}}(\tau).
\label{Equation_3}
\end{eqnarray} 
Next, ${\cal{S}}_{B}[\bar{f},f]$ and ${\cal{S}}_{B}[\bar{c},c]$ are Berry actions for $f$ and $c$-electrons and they are defined as follows ${\cal{S}}_{B}[\bar{x},x]=\sum_{{\bf{r}}}\int^{\beta}_{0}d\tau \bar{x}({\bf{r}}\tau)\dot{x}({\bf{r}}\tau)$, where $\dot{x}({\bf{r}}\tau)=\partial_{\tau}x({\bf{r}}\tau)$ is the time derivative. 
We perform the local gauge transformation to new fermionic variables $\tilde{f}({\bf{r}}\tau)$ and $\tilde{c}({\bf{r}}\tau)$. For the electrons of $f$ and $c$ orbitals, the U$(1)$ gauge transformation could be written as
\begin{eqnarray}
\left[
\begin{array}{cc}
x({\bf{r}}\tau) \\
\bar{x}({\bf{r}}\tau)
\end{array}
\right]=\hat{{\cal{U}}}(\varphi)\left[
\begin{array}{cc}
\tilde{x}({\bf{r}}\tau) \\
\bar{\tilde{x}}({\bf{r}}\tau)
\end{array}
\right],
\label{Equation_4}
\end{eqnarray} 
where $\hat{\cal{U}}(\varphi)$ is the U(1) transformation matrix $\hat{\cal{U}}(\varphi)=\hat{I}\cos\varphi({\bf{r}}\tau)+i\hat{\sigma}_{z}\sin\varphi({\bf{r}}\tau)$, where $\varphi({\bf{r}}\tau)$ are the new phase variables, $\hat{I}$ is the unit matrix, and $\hat{\sigma}_{z}$ is the $z$ component of the Pauli matrix. Then, the decoupling of the nonlinear density terms in the action is rather standard, and we do not present here the calculation details. We give only the final form of the total action of the system in the Fourier-space after the transformation in Eq.(\ref{Equation_4}) and the linearization procedure.  
\begin{eqnarray}
&&{\cal{S}}_{\rm eff}\left[\tilde{\bar{c}},\tilde{c},\tilde{\bar{f}},\tilde{f}\right]
=\frac{1}{\beta{N}}\sum_{{\bf{k}}\nu_{n}}\left[\bar{\tilde{c}}_{\bf{k}}(\nu_{n}),\bar{\tilde{f}}_{\bf{k}}({\nu_{n}})\right]\times
\nonumber\\
&&\times{\cal{G}}^{-1}({\bf{k}},\nu_{n})\left[\begin{array}{cc}
{\tilde{c}}_{\bf{k}}(\nu_{n})\\
{\tilde{f}}_{\bf{k}}(\nu_{n})
\end{array}
\right].
\label{Equation_6}
\end{eqnarray}
Here $\nu_{n}={\pi(2n+1)/\beta}$, $n=0,\pm1,\pm2, ...$ are fermionic Matsubara frequencies, and  ${\cal{G}}^{-1}({\bf{k}},\nu_{n})$ is the inverse of the Green function matrix, given by
\begin{eqnarray}
{\cal{G}}^{-1}({\bf{k}},\nu_{n})=
\left(
\begin{array}{cc}
{E}^{\tilde{c}}_{{\bf{k}}}(\nu_{n})
 & -\bar{\Delta}  \\
-\Delta & {E}^{\tilde{f}}_{{\bf{k}}}(\nu_{n})
\end{array}
\right),
\label{Equation_7}
\end{eqnarray}
where single-particle Bogoliubov's quasienergies ${E}^{\tilde{f}}_{{\bf{k}}}(\nu_{n})$ and ${E}^{\tilde{c}}_{{\bf{k}}}(\nu_{n})$ are given as ${E}^{\tilde{c}}_{{\bf{k}}}(\nu_{n})=\bar{\epsilon}_{\tilde{c}}-i\nu_{n}-{t}_{{\bf{k}}}$, ${E}^{\tilde{f}}_{{\bf{k}}}(\nu_{n})=\bar{\epsilon}_{\tilde{f}}-i\nu_{n}-\tilde{t}_{{\bf{k}}}$. Next, ${t}_{{\bf{k}}}$ and $\tilde{t}_{{\bf{k}}}$ are band-renormalized hopping amplitudes ${t}_{{\bf{k}}}=2t{\mathrm{g}}_{B}\gamma_{{\bf{k}}}$ and $\tilde{t}_{{\bf{k}}}=2\tilde{t}{\mathrm{g}}_{B}\gamma_{{\bf{k}}}$, where ${\mathrm{g}}_{B}$ is the bandwidth renormalization factor 
$\mathrm{g}_{B}=\left.\left\langle e^{-i[\varphi({{\bf{r}}}\tau)-\varphi({{\bf{r}}}'\tau)]} \right\rangle\right|_{|{\bf{r}}-{\bf{r}}'|={{a}}}
$ and $\gamma_{{\bf{k}}}$ is the 3D lattice dispersion $\gamma_{{\bf{k}}}=\cos(k_{x})+\cos(k_{y})+\cos(k_{z})$. The quasiparticle energies $\bar{\epsilon}_{\tilde{f}}$ and $\bar{\epsilon}_{\tilde{c}}$ are of Hartree-type and they are defined in the theory by relation $\bar{\epsilon}_{\tilde{x}}=\epsilon_{{x}}-\mu+Un_{\tilde{y}}+i\left\langle\dot{\varphi}({{\bf{r}}}\tau)\right\rangle$, where $\tilde{y}$ means orbital, opposite to $\tilde{x}$.

\begin{figure}[!ht]
\includegraphics[width=160px,height=160px]{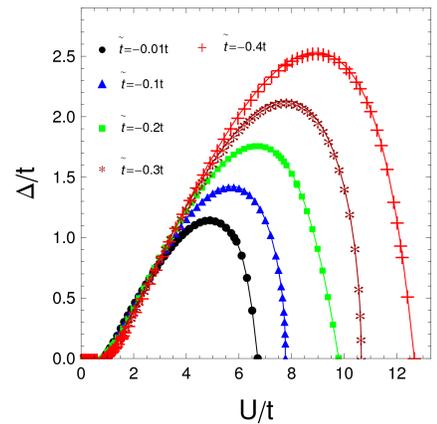} 
{\caption{\label{fig:Fig_1} The local excitonic order parameter $\Delta$ normalized to the $c$-band hopping amplitude $t$ as a function of the on-site Coulomb interaction parameter $U/t$. Different values of $\tilde{t}$ are considered.}}
\end{figure}

We get a set of coupled self-consistent equations for the EI order parameter $\Delta$
\begin{eqnarray}
&&\frac{1}{N}\sum_{{\bf{k}}}\left[f({E}^{+}_{{\bf{k}}})+f({E}^{-}_{{\bf{k}}})\right]=1,
\label{Equation_8} 
\newline\\
&&\tilde{n}=\frac{1}{N}\sum_{{\bf{k}}}\xi_{{\bf{k}}}\frac{f({E}^{+}_{{\bf{k}}})-f({E}^{-}_{{\bf{k}}})}{\sqrt{\xi^{2}_{{\bf{k}}}+4\Delta^{2}}},
\label{Equation_9} 
\newline\\
&&\Delta=-\frac{U\Delta}{N}\sum_{{\bf{k}}}\frac{f({E}^{+}_{{\bf{k}}})-f({E}^{-}_{{\bf{k}}})}{\sqrt{\xi^{2}_{{\bf{k}}}+4\Delta^{2}}}.
\label{Equation_10}  
\end{eqnarray}
Here $N$ is the total number of lattice sites, $\xi_{{\bf{k}}}=-{t}_{{\bf{k}}}+\bar{\epsilon}_{\tilde{c}}+\tilde{t}_{{\bf{k}}}-\bar{\epsilon}_{\tilde{f}}$ is the quasiparticle dispersion and the energy parameters ${E}^{+}_{{\bf{k}}}$ and ${E}^{-}_{{\bf{k}}}$ are defined as
\begin{eqnarray}
{E}^{\pm}_{{\bf{k}}}=\frac{1}{2}\left(-{t}_{{\bf{k}}}+\bar{\epsilon}_{\tilde{c}}-\tilde{t}_{{\bf{k}}}+\bar{\epsilon}_{\tilde{f}}\pm{\sqrt{\xi^{2}_{{\bf{k}}}+4\Delta^{2}}}\right).
\label{Equation_11}
\end{eqnarray}
In Fig.~\ref{fig:Fig_1} the numerical results for the local excitonic order parameter $\Delta$ are presented. The region, where $\Delta\neq 0$ defines the EI phase in the system \cite{cite-11, cite-12,cite-13,cite-14}.
%
%
\section{\label{sec:Section_4} Phase stiffness and condensation}
\begin{figure}[!ht]
\includegraphics[width=160px,height=160px]{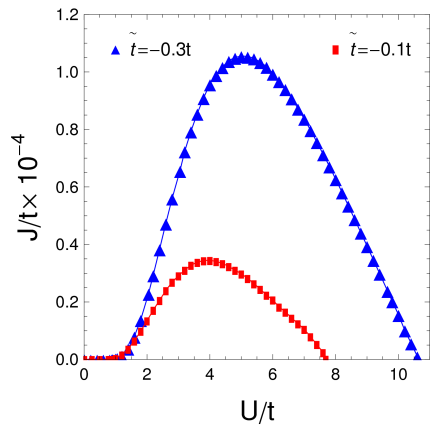} 
{\caption{\label{fig:Fig_2} The excitonic phase stiffness parameter $J$ given in Eq.(\ref{Equation_16}) as a function of the on-site Coulomb interaction parameter $U/t$. Two different values of the $f$-band hopping amplitude are considered.} }
\end{figure}

%
In this Section we integrate out the fermions in the partition function in Eq.(\ref{Equation_2}) and we obtain the bosonic total phase action of the system. We will show how the \textit{non-local} fermionic correlations give the main contribution to the phase stiffness of the ensemble of excitons. The partition function in Eq.(\ref{Equation_2}) could be rewritten as 
\begin{eqnarray}
{\cal{Z}}=\int\left[{\cal{D}}\varphi\right] e^{-{\cal{S}}_{\rm eff}[\varphi]},
\label{Equation_12}
\end{eqnarray}
where the effective phase action in the exponential is ${\cal{S}}_{\rm eff}[\varphi]=-\ln\int\left[{\cal{D}}\bar{\tilde{c}}{\cal{D}}\tilde{c}\right]\left[{\cal{D}}\bar{\tilde{f}}{\cal{D}}\tilde{f}\right] e^{-{{\cal{S}}}[\bar{\tilde{c}},{\tilde{c}},\bar{\tilde{f}},{\tilde{f}},\varphi]}$. After expanding the logarithm up to second order in the cumulant series expansion (higher terms are not considered), we find for the important part of the effective phase action
\begin{eqnarray}
{\cal{S}}_{\rm eff}[\varphi]={\cal{S}}_{0}[\varphi]+{\cal{S}}_{J}[\varphi],
\label{Equation_13}
\end{eqnarray} 
where ${\cal{S}}_{0}[\varphi]$ is the phase-only action after U(1) gauge transformation 
\begin{eqnarray}
{\cal{S}}_{0}[\varphi]=\sum_{{\bf{r}}}\int^{\beta}_{0}d\tau\left[\frac{\dot{\varphi}^{2}({\bf{r}}\tau)}{U}-\frac{2\bar{\mu}}{iU}\dot{\varphi}({\bf{r}}\tau)\right]
\label{Equation_14}
\end{eqnarray}  
and ${\cal{S}}_{J}[\varphi]=-\frac{1}{2}\left\langle{\cal{S}}^{2}\right\rangle_{{\cal{S}}_{\rm eff}}$. After calculating all averages in the expression of ${\cal{S}}_{J}[\varphi]$ and after not complicated evaluations we rewrite the action ${\cal{S}}_{J}[\varphi]$ in the form
\begin{eqnarray}
{\cal{S}}_{J}\left[\varphi\right]=-\frac{J}{2}\int^{\beta}_{0}d\tau \sum_{\left\langle{\bf{r}},{\bf{r}}'\right\rangle}\cos{2\left[\varphi({\bf{r}}\tau)-\varphi({\bf{r}}'\tau)\right]},
\label{Equation_15}
\end{eqnarray}
where the exciton phase stiffness parameter $J$ is given by the relation
\begin{eqnarray}
J=&&\frac{\Delta^{2}t\tilde{t}}{{N^{2}}}\sum_{{\bf{k}},{\bf{k}}'}\frac{\gamma_{{\bf{k}}}\gamma_{{\bf{k}}'}}{{\sqrt{\xi^{2}_{{\bf{k}}}+4\Delta^{2}}}}\left[\Lambda_{1}({\bf{k}},{\bf{k}}')\tanh\left(\frac{\beta {E}^{+}_{{\bf{k}}}}{2}\right)-\right.
\nonumber\\
&&\left.-\Lambda_{2}({\bf{k}},{\bf{k}}')\tanh\left(\frac{\beta {E}^{-}_{{\bf{k}}}}{2}\right)\right].
\label{Equation_16}
\end{eqnarray}
The parameters $\Lambda_{1}({\bf{k}},{\bf{k}}')$ and $\Lambda_{2}({\bf{k}},{\bf{k}}')$ in Eq.(\ref{Equation_16}) are defined as
\begin{eqnarray}
\Lambda^{-1}_{1,2}({\bf{k}},{\bf{k}}')=\left({{E}^{\pm}_{{\bf{k}}} - {E}^{\pm}_{{\bf{k}}'}}\right)\left({E^{\pm}_{{\bf{k}}} - {E}^{\mp}_{{\bf{k}}'}}\right)
\end{eqnarray}
The form of $J$ in Eq.(\ref{Equation_16}) indicates that the phase stiffness in the system of excitonic pairs is characterized by an energy scale proportional to $(\Delta t_{e}t_{h})/({t_{e}+t_{h}})$ for all the values of the Coulomb interaction parameter $U$ and it is related to the motion of the center of mass of e-h composed quasiparticle, because $(t_{e}t_{h})/(t_{e}+t_{h}) \approx (m_{e}+m_{h})^{-1}$, \cite{cite-9} implying
that the exchange coupling parameter becomes proportional to the excitonic BEC critical temperature \cite{cite-9, cite-19}. The numerical evaluations of $J$ for the case $T=0$ are shown in Fig.~\ref{fig:Fig_2}.
%
%

\section{\label{sec:Section_5} Coherent DOS spectra}

We define here the $c$ and $f$ -band normal single-particle excitonic Green functions $G_{\rm x,x}({\bf{r}}\tau,{\bf{r}}'\tau')=-\langle {x}({\bf{r}}\tau)\bar{x}({\bf{r}}'\tau')\rangle$. After introducing the U(1) transformations, defined in Eq.(\ref{Equation_4}), we will have the Green function's decomposition into two parts: purely fermionic and bosonic correlation function: $G_{\rm x,x}({\bf{r}}\tau,{\bf{r}}'\tau')=-\langle \tilde{x}({\bf{r}}\tau)\bar{\tilde{x}}({\bf{r}}'\tau')\rangle\langle e^{-i\left[\varphi({\bf{r}}\tau)-\varphi({\bf{r}}'\tau')\right]}\rangle,
$. For the fermionic correlation function we get
\begin{eqnarray}
\tilde{G}_{\rm \tilde{x},\tilde{x}}\left({\bf{k}},i\nu_{n}\right)=\frac{{E}^{\tilde{y}}_{{\bf{k}}}\left(\nu_{n}\right)}{E^{\tilde{x}}_{{\bf{k}}}\left(\nu_{n}\right){E}^{\tilde{y}}_{{\bf{k}}}\left(\nu_{n}\right)-\Delta^{2}}.
\label{Equation_17}
\end{eqnarray}
%
%
\begin{figure}[!ht]
\includegraphics[width=200px,height=200px]{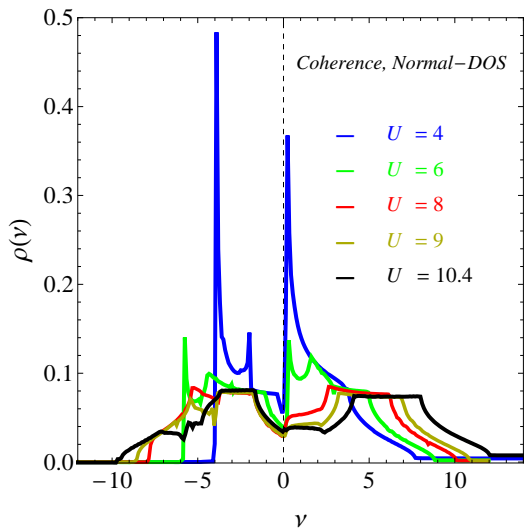} 
{\caption{\label{fig:Fig_3} The total single-particle coherent DOS function given in Eq.(\ref{Equation_19}.) Different values of the Coulomb interaction parameter are considered.} }
\end{figure}
%
Indeed, the single-particle density of states is related with the imaginary part of the retarded Green functions $\rho_{\rm \tilde{x},\tilde{x}}\left({\bf{k}},\nu\right)=-\frac{1}{\pi}\Im\tilde{G}^{R}_{\rm x,x}({\bf{k}},\nu)$, thus we need to calculate real retarded function, which corresponds to the normal Matsubara Green function $\tilde{G}_{\rm \tilde{x},\tilde{x}}\left({\bf{k}},i\nu_{n}\right)$. This could be done by the analytical continuation into the upper-half complex semi-plane ($\nu_{n}>0$) of frequency modes $i \nu_{n}$
\begin{eqnarray}
\tilde{G}^{R}_{\rm \tilde{x},\tilde{x}}({\bf{k}},\nu)=\tilde{G}_{\rm \tilde{x},\tilde{ x}}\left({\bf{k}},i\nu_{n}\right)|_{i\nu_{n}\rightarrow \nu+i\eta}.
\label{Equation_18}
\end{eqnarray}
The nonlocal phase-phase correlation function could be calculated in the frame of the quantum rotor phase action discussed in the Section \ref{sec:Section_4}. We do not present here the calculation details and we give only the final form of the single-particle excitonic normal DOS function
\begin{eqnarray}
&&\rho_{\rm x,x}(\nu)=|\psi_{0}|^{2}\rho_{\rm \tilde{x},\tilde{x}}(\nu)-
\nonumber\\
&&-U\int^{+3}_{-3}dx \frac{\rho_{3D}(x)}{4\sqrt{\bar{\mu}^{2}+4UJ\left(3-x\right)}}\times
\nonumber\\
&&\left\{\rho_{\rm \tilde{x},\tilde{x}}\left(\nu-\kappa_{1}\left(x\right)\right)\left[n\left(\kappa_{1}(x)\right)+f\left(\nu-\kappa_{1}(x)\right)\right]+\right.
\nonumber\\
&&\left.\rho_{\rm \tilde{x},\tilde{x}}\left(\nu-\kappa_{2}\left(x\right)\right)\left[n\left(\kappa_{2}(x)\right)+f\left(\nu-\kappa_{2}(x)\right)\right]\right\},
\label{Equation_19}
\end{eqnarray}
where $|\psi_{0}|^{2}$ is the BEC transition probability function, and the functions $\rho_{\rm \tilde{x},\tilde{x}}\left(\nu\right)$ in Eq.(\ref{Equation_19}) are given by 
\begin{eqnarray} 
\rho_{\rm \tilde{x},\tilde{x}}\left(\nu\right)=\int^{+3}_{-3}dx \rho_{ 3D}(x)\frac{\left[\bar{\epsilon}_{\tilde{y}}-\tilde{t}(x)-\nu\right]^{2}}{\sqrt{\xi^{2}(x)+4\Delta^{2}}}\times
\nonumber\\
\times\left\{\frac{\delta\left[\nu-{E}^{+}(x)\right]}{|\bar{\epsilon}_{\tilde{y}}-\tilde{t}(x)-{E}^{+}(x)|}+\frac{\delta\left[\nu-{E}^{-}(x)\right]}{|\bar{\epsilon}_{\tilde{y}}-\tilde{t}(x)-{E}^{-}(x)|}\right\}.
\label{Equation_20}
\end{eqnarray}

The function $\rho_{3D}(x)$ in Eqs.(\ref{Equation_19}) and (\ref{Equation_20}) is the DOS function for the 3D cubic lattice $\rho_{3D}(x)=\frac{1}{N}\sum_{{\bf{k}}}\delta(x-\gamma_{\bf{k}})$. The functions $n\left(x\right)$ and $f\left(x\right)$ in Eq.(\ref{Equation_19}) are the Bose-Einstein and Fermi-Dirac distribution functions respectively. The parameters in the arguments of distribution functions, are $\kappa_{1,2}(x)=-\bar{\mu}\pm{\sqrt{\bar{\mu}^{2}+4U{J}\left(3-x\right)}}$.
The presence of singularities in the integration
region in Eq.(\ref{Equation_20}) causes that we used an adaptive 21-point integration routine combined with
the Wynn $\epsilon$-algorithm \cite{cite-20} to calculate those integrals numerically. The accuracy for
adaptive evaluations is achieved with a relative error of order of $10^{-7}$.
The resulting coherent DOS spectra is given in Fig.~\ref{fig:Fig_3}, where we see how the strong coherence effects in the excitonic system suppress the hybridization gap \cite{cite-11,cite-12,cite-13,cite-14} in the single particle DOS spectra. The gapless behavior of the single-particle DOS is a direct consequence of the phase-stiffness mechanism of the excitonic condensation. 
%
\section{\label{sec:Section_6} Final remarks}

We have considered the problem of excitonic condensation within a two band solid state model. The EI state is derived in the form of the local excitonic gap parameter. Then, considering the bosonic phase sector, we have derived the excitonic phase stiffness parameter, which we found as responsible for the excitonic condensation mechanism. The form of it suggests that it is related to the motion of the center of mass of the e-h quasiparticle, thus implying the relation with the coherent excitonic condensate state. Furthermore, we have calculated the coherent single-particle normal DOS spectra for different values of the Coulomb interaction parameter. We have shown that there is no hybridization gap in the system, and the DOS spectra is gapless for all values of the Coulomb interaction parameter $U$. 
%
\section*{References}
%

\end{document}